\documentclass[twocolumn]{article}
\usepackage[T1]{fontenc}
\usepackage{graphicx}
\usepackage{epsfig}
\usepackage[latin9]{inputenc}
\usepackage{listings}
\begin{document}
\bibliographystyle{plain}

\thispagestyle{plain}

\title{A Generalized Streaming Model For Concurrent Computing\footnote{First draft: Jan. 31, 2010; this draft: Dec. 07, 2010.}}

\author{Yibing Wang\footnote{Author contact: williamwang@acm.org.}}

\date{}
\maketitle
{\small
\parskip=0pt
\parindent=0pt

{\bf Abstract-}
Multicore parallel programming has some very difficult problems
such as deadlocks during synchronizations and race conditions 
brought by concurrency. Added to the difficulty is the lack of
a simple, well-accepted computing model for multicore 
architectures---because of that it is hard to develop powerful 
programming environments and debugging tools. To tackle the
challenges, we promote a generalized stream computing model,
inspired by previous researches on stream computing, that unifies 
parallelization strategies for programming language design, 
compiler design and operating system design. Our model provides 
a high-level abstraction in designing language constructs to 
convey concepts of concurrent operations, in organizing a program's 
runtime layout for parallel execution, and in scheduling 
concurrent instruction blocks through runtime and/or operating 
systems. In this paper, we give a high-level description of the proposed 
model: we define the foundation of the model, show its simplicity 
through algebraic/computational operation analysis, illustrate a 
programming framework enabled by the model, and demonstrate its 
potential through powerful design options for programming 
languages, compilers and operating systems.
}

\section{Introduction}
\pagestyle{plain}

\subsection{Motivation}
When searching for a general approach to the challenges~\cite{berkeleyview}
faced by multicore programming, particularly deadlocks during 
synchronizations and race conditions~\cite{races} introduced by
concurrency, which render solutions built on conventional multithreading 
models~\cite{posixthread,openmp,c++ext} unappealing to programmers,
we found that there was a need for a unified computing model on which
computer professionals at programming language design level,
compiler design level, and operating system design level could benefit
from each others' works; and a generalization of the stream computing
model~\cite{buck2006} seemed to have the potential to serve as such a 
design model for future generation computing. Some reasons are explained 
as follows.

Information processing on modern computers has never followed a monotonous
path, but the mainstream has come along a relatively unified way due
to the Turing machine abstraction in theory~\cite{hopcroft} and von Neumann
model~\cite{backus1978} in design. The adoption of multicore architectures,
however, brings the industry into
a new chaos at various computing levels ranging from programming
language designs to operating system designs. Not only are programmers
often frustrated with primitive solutions or the lack of ideal software
tools (e.g. programming languages and debuggers), but also are chip makers 
facing challenges in delivering new designs~\cite{larrab,larrabee}; whereas
tools' builders are constrained by the multithreading model, such as in the 
cases of OpenMP~\cite{openmp} and C++-0x thread library extension~\cite{c++ext},
or by imperative languages, such as in the cases of CUDA~\cite{nick,nvidia}
and OpenCL~\cite{opencl}. To change the impasse situation where researchers
and developers at different computing levels have uncertain
expectations from each other, first a well-accepted computing model
is needed, i.e. we need a model that can guide designs at
different computing levels. 

PRAM~\cite{prambook} has often been used as a conceptual parallel 
computing model in demonstrating synchronized executions such as 
concurrent read concurrent write and concurrent read exclusive write. 
We argue that PRAM captures certain essences such as parallelism in
concurrent computing\footnote{In this paper, \textit{concurrent} and
\textit{parallel} are interchangeable in most cases, though we tend to regard
concurrent as having broader meaning in computing than parallel does.
The authors of~\cite{cleaveland96} think parallel systems are mostly
synchronous, whereas concurrent systems also include distributed cases
that are mostly asynchronous.} but not other substances such as
precedence constraints in operation; and the very nature of its
dependence on shared memory limits PRAM's power in design. Therefore,
if PRAM is only a generalization (or parallelization) of the RAM~\cite{hopcroft}
model (Turing machine equivalent) for \emph{abstract analysis}, what
is the parallel-formed generalization of the von Neumann model and
its alternatives for \emph{design}? Conventional threads and transactional
memory~\cite{tm1,tm2,tm3} are not the answer; they are implementation
details with little extension to the von Neumann model. Many people
agree that threads are low-level programming utilities and should
remain so; transactional memory may not be computationally 
productive nor energy-efficient when transactions are large in 
size or long in time~\cite{tm2}.

Stream computing~\cite{buck2006,dally2003,labonte2004,scout2004} has
been around for more than a decade. Perhaps due to their humble origin
(i.e. graphics processors; not their noble originators) in the computer
industry, technologies based on stream computing were often used only
as performance accelerators. While surveying the landscape of multicore
parallel computing, we realized that, maybe, the application of stream
computing model should go beyond even GPGPU computing, and a 
generalization of the model could be applied to general purpose software
designs and programming as well. Note our pursuit of a general design
model was not to find a killer solution (i.e. the implementation
of a certain method) for all applications, but to find a design principle
in handling some of the most difficult yet fundamental problems, such as
parallelization and nondeterminism.

To serve our purpose, we recommends the separation of computing models
for \emph{abstract analysis} from computing models for \emph{generalized design}. In
\cite{snyder2007}, Snyder expressed similar idea but from a different
angle; he emphasized that a useful parallel machine model should
be able to capture key features that have impact on performance,
which to us is a design (model) issue. Based on such understanding,
plus other people's publications~\cite{threadlib,tm1,gumm,herlihy,tm3,ealee1,ealee2,liao}
and our own experiences\footnote{For example, the author wrote his
first multithreaded server in 1997. Part of that early experience was
published in~\cite{wang2000}.}, we propose a generalized stream
computing (hereinafter known as GeneSC, pronounced {}``g-e-n-e-s-i-s'') 
model as a guideline for multicore computing. Our expectations of 
the model can be summarized as follows.
\begin{itemize}
\item It will help with defining language constructs for expressing 
concurrency structures in programs and algorithms, but leaving parallelization
and synchronization controls to compilers, runtimes and/or operating
systems.
\item It will bring parallelizing compilers out of the confinement of 
sequential execution constructs such as loop nests.
\item It will bring in operating system designs for tractable, reliable
and concurrent instruction sets executions.
\item It will support backward compatible programming schemes, be it 
structured, object-oriented, or functional; and the re-use of existing
(modified when necessary) library routines.
\item It can be reduced to sequential case for debugging purpose or
when used for single core architectures.
\end{itemize}

Furthermore, we regard algorithm design as an inseparable part of achieving 
high-performance in multicore parallel computing. We believe that, due to
different algorithmic thinking, solutions to an application
problem may show different concurrency structures, which have different
levels of difficulty when mapped to the underlying runtime environment
and hardware. By supplying an expressive computing model, we give
programmers the needed power in defining the lest convoluted concurrency
structures in their solutions. We use the following example to 
further justify our motivation.

\subsection{Example}

A naive solution to the $N$-body problem~\cite{barneshut,wilkinson} has 
the computation complexity of \(O(N^{2})\). Figure~\ref{fig:nbody1} shows
such an algorithm, where gforce() is a function for calculating 
gravitational forces from the other $N-1$ objects to the $j$th object,
following the Newtonian laws of physics. To parallelize this algorithm,
we often rely on loop nest parallelization by hand or through compilers~\cite{allen,gerber}.

\begin{figure}[h]
\small
\inputencoding{latin1}
\begin{lstlisting}[basicstyle={\footnotesize}]
  1 /* calculate gravitational forces in time
  2    sequence from zero to tmax */
  3 for (ti = 0; ti < tmax; ti++)
  4 {
  5    for (j = 0; j < N; j++)
  6    {
  7       f = gforce(j);
  8       vnew[j] = v[j] + f * dt;
  9       xnew[j] = x[j] + vnew[j] * dt;
 10    }
 11    for (j = 0; j < N; j++)
 12    {
 13       v[j] = vnew[j];
 14       x[j] = xnew[j];
 15    }
 16 }
\end{lstlisting}
\caption{\small The $N$-body algorithm of complexity \(O(N^{2})\).} 
\label{fig:nbody1}
\end{figure}

Instead of doing \(O(N^{2})\) direct force integrations, an improved
design~\cite{barneshut} recursively divides the space into
smaller cubic cells until each cubic cell contains no or only one of
the $N$ objects, then builds a tree with cubic cells that have more
than one objects as the intermediate nodes and cells with only one
object as the leaves. Gravitational forces among the $N$ objects are
calculated in a way where far away objects are approximated by their
conglomeration masses at their geometry centers. Figure~\ref{fig:nbody2}
shows the stream-like skeleton of such an algorithm with complexity 
\(O(Nlog^{N})\), where each inner-loop function consists data parallelism.

\begin{figure}[h]
\small
\inputencoding{latin1}
\begin{lstlisting}[basicstyle={\footnotesize}]
  1 for (ti = 0; ti < tmax; ti++)
  2 {
  3    space_subdivision();
  4    tree_construction();
  5    mass_center_calc();
  6    approximate_force();
  7    position_update();
  8 }
\end{lstlisting}
\caption{\small An improved $N$-body algorithm of complexity \(O(Nlog^{N})\).}
\label{fig:nbody2}
\end{figure}

The $N$-body problem is interesting in three ways. First, it is a large
computation problem that must be solved on parallel computers. Second,
inter-node communication (i.e. synchronization) overhead in executions 
of each parallelized inner functions is not negligible. Third, it has 
the potential to show how languages (i.e. the expression of reasoning) 
may affect algorithm designs.

Further improved algorithms such as~\cite{bhatt92,warren92} aim to reduce 
communication overhead. Multicore processors may reduce the overhead even
more through shared memory. Besides hardware support, a suitable
computing model is needed. The GeneSC model that
we propose has considered such applications as well as many others.

\subsection{Organization}
The rest of the paper explains what is this generalized stream computing
and what are the enabling technologies. Section~\ref{sec-define}
first highlights a few important concepts in the problem domain of interests
and then gives a definition of the GeneSC model, followed
by Section~\ref{sec-oper} on algebraic/computational operation analyses,
Section~\ref{sec-prog} on a programming framework enabled by our model,
and Section~\ref{sec-design} on design considerations including three
crucial additions that depart from traditional practices. Section~\ref{sec-app}
introduces one potential application. Section~\ref{sec-related} discusses
related works. Section~\ref{sec-conclude} gives the concluding remarks.

\section{Definition}
\label{sec-define}

We use synchronism and asynchronism for discussing operations' timing
or ordering characteristics. We use determinism and non-determinism
for discussing bounded program behaviors, with non-determinism
refers to program behaviors that are different (but well-defined)
from run to run even for the same given inputs; and indeterminacy
for unbounded, i.e. unpredictable program behaviors. Synchronous
operations without randomized functions normally yield
deterministic results; asynchronous operations may yield either
deterministic (when no ordering is needed) or non-deterministic
(when ordering is required but, by mistake, not enforced) results. Shared
memory could add more complexities, e.g. when supposedly deterministic
operations show non-deterministic or even indeterminacy behaviors,
or when non-deterministic operations show indeterminacy behaviors, due
to race conditions.

Deadlocks and race conditions are notorious because they bring
indeterminacy. General races~\cite{races} can be defined as 
concurrent accesses to shared data with at least one access 
is a write operation. General races may be acceptable if they
just cause non-determinism, but are unacceptable if they cause
indeterminacy. Data races are special cases of general
races; data races are program bugs, e.g. missing shared locks
in concurrent accesses to shared data.

Applications show parallelism in two ways: data parallelism and task 
parallelism (we consider pipelined concurrency as a subtype of task
parallelism). Performance in data parallelism cases relies on data 
processing speed, whereas performance in task parallelism cases relies
on overall throughput.

To make a parallel program work correctly, we prevent indeterminacy;
to make the behavior of a parallel program tractable, we deal with
non-determinism.

In our GeneSC model, programming, compilation 
and runtime designs adopt a concept that views computing as applying
well-defined functions to flows of data sets, which we call 
\textit{stream data}. At the core of the model is an entity, 
\textit{E}\,=\,(\textit{k},\,\textit{d},\,\textit{r}), that encapsulates an
execution unit consisting of three elements:
\begin{itemize} 
\item \textit{k}: a close-form function
\item \textit{d}: well-defined input/output data
\item \textit{r}: relations with other entities
\end{itemize}
The close-form function is in fact a non-trivial private algorithm
and is often called a (mathematical) kernel. The input/output
data defines a function's interactions with the outside, without
exposing intermediate computation results. Inter-stream-entity relations
offer timing/ordering dependence.

This model is different from conventional multithreading models in at
least two ways: First, stream entities are functional units, whereas
threads are execution units. Second, explicit concurrency structures
defined by stream entity's syntax and semantics allow optimization in
compilation and scheduling at runtime, whereas threads prescribe
execution structures without rich timing/ordering knowledge.

Some people might suggest that a stream entity resembles a neuron
in artificial neuron networks, except that the former normally has a
linear function whereas the latter a non-linear function. We will not
diverge too far along that direction.

\section{Operations}
\label{sec-oper}

\begin{figure}[h]
{\centering \resizebox*{0.45\textwidth}{!}{\includegraphics{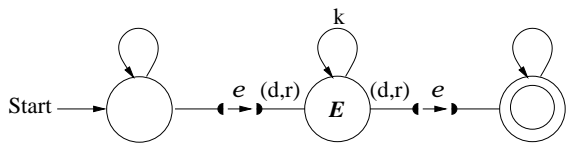}} \par}

\caption{\small An abstract operation model of a program represented by a finite automata with $\epsilon$-moves in the proposed streaming model nomenclature.}
\label{fig:auto}
\end{figure}

According to its definition, a stream entity contains a miniature
Turing machine whose input symbols are from a subset or a segment 
of all input symbols to a larger Turing machine, and whose finite
control is defined by the stream entity's kernel function. But
the overall program defined by stream entities operates as a 
finite automata (FA), where each state in the FA has a miniature
Turing machine inside. Inter-stream-entity relations define precedence 
constraints among the FA states but do not define connection 
(i.e. transition) policies, thus the program operates as a FA with 
$\epsilon$-moves, shown in Figure~\ref{fig:auto}, where the
$\epsilon$-moves simply mean undefined transition functions, not 
important to stream entities\footnote{While reading Snyder's
paper~\cite{snyder2007} after the conception of the GeneSC
model, the author could instantly recognize the similarity between his
ideal of an operational stream computing model and Snyder's
considerations for the candidate type architecture (CTA) abstraction
for MIMD parallel machines.

As another note, from our point of view, one major difference between
multicore and manycore resembles the difference between shared-memory
systems and distributed memory (i.e. networking) systems, where
inter-process communication mechanisms set them apart. Our GeneSC
model supports both systems.}. 

The GeneSC model has all the advantages such as concise
foundations and simple semantics of simple operational models
without being weak in program clarity, which is guaranteed by
detailed designs and implementations of individual stream entities.
For examples, addition operations in the form of parallel executions
of stream entities are supported; concatenation operations are 
allowed as long as inter-stream-entity relations, if any, are preserved;
also allowed are composition operations. With 
these operations, building large-scale software systems and reusing
library components are safe and tractable. The composition capability
is even more crucial for hierarchical and scalable computing, which 
will be discussed in Sections~\ref{sec-compiler} and \ref{sec-os}. 

The proposed streaming model preserves asynchronism inherited from the
problem domain, but reduces non-determinism to the minimum  through
explicit concurrency structures in a program\footnote{This
model provides the benefit of functional programming in compositional
operations but is also intended to yield better performance.}. By
encapsulating internal computation states, isolating external 
states, and separating computation from communication, stream
entities help reducing data races. Meanwhile, inter-stream-entity relations
help reducing general races.

\section{Programming}
\label{sec-prog}

\begin{figure}[h]
{\centering \resizebox*{0.45\textwidth}{!}{\includegraphics{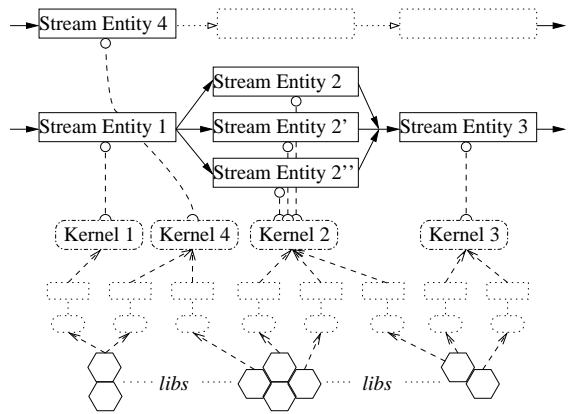}} \par}

\caption{\small A programming framework based on the GeneSC model.}
\label{fig:prog}
\end{figure}

Figure~\ref{fig:prog} shows a programming framework based on the GeneSC
model. At the top level, programmers focus on coding 
concurrency structures defined by stream entities, which relate to each other 
according to precedence constraints, if any, that exist in high-level control
flows and data flows. Each stream entity has a kernel function, which may 
consist of none, one or more than one lower-level stream entities. The lowest 
level kernel functions will call routines in conventional (e.g. non-threaded
but thread-safe) programming libraries.

In the above figure, hierarchical structures among stream entities embody
compositional operations. Dependence constraints among stream entities 
decide their runtime execution orders; if there is no dependence constraint
between two stream entities, there is no mandatory execution order between
them. For example, because stream entity No.4 is independent of others,
it may be executed in parallel with stream entity No.1, No.2 or No.3.
Programmers see no thread or any low-level parallel execution controls,
which are left to compilers, runtime and operating systems to handle.
Programmers do have the responsibility to write kernel functions and
conventional library routines, if there are no existing ones for reuse.

\section{Designs}
\label{sec-design}

The GeneSC model provides a highly abstracted, mostly 
tractable view of concurrent operations. Here are some \textit{examples} 
as design considerations that support the programming framework discussed
in Section~\ref{sec-prog}.

\subsection{Language}
\label{sec-lang}

Stream entity is the foundation of conveying concurrency structures in
a program. Though similar concept may already exist in some programming
languages, we decide to make it explicit. A stream entity's API may
have the following contents in a new programming-language construct:

\begingroup
\small
\inputencoding{latin1}
\begin{lstlisting}[basicstyle={\footnotesize}]
  {
    relations:
    {
      before:{(function_k,[hard/soft]), ...}
      after: {(function_i,[hard/soft]), ...}
    }
    input:  <T> x
    output: <T> y
    kernel: function_j
  }
\end{lstlisting}
\endgroup

Inter-stream-entity relations are defined by two sets. Those sets can be empty. 
Elements in the relation sets are kernel function and relation constraint 
pairs. A {}``hard'' relation constraint means {}``this'' entity \emph{must} finish 
\emph{before} related entities begin, or \emph{must} begin \emph{after} 
related entities finish; a {}``soft'' constraint means {}``this'' entity
\emph{should} finish \emph{before} related entities begin, or {}``this'' 
entity \emph{should} begin \emph{after} related entities finish.
Compositional relations are generated automatically by compilers and are 
kept in an internal data structure for stream entities.

One major difference between a stream entity and a conventional task
in the form of a thread is that the former makes composition operations
easier and safe, whereas the latter harder and unsafe. One major difference
between a stream entity and a class object is that the former performs
an overall well-defined, close-form function independent of context,
whereas the latter does not perform such a function without context, except
in the case of a function object or functor~\cite{c++}. Different from a
functor, a stream entity's internal states are not important.

Stream entity does not imply how its kernel function is
implemented, thus does not enforce an imperative language implementation
targeting CPU or GPU, or a functional language implementation, or
something else. Building stream entities on existing, low-level
library routines is possible provided that the library routines do not
create POSIX-like threads. In the proposed streaming model, user 
programs are not encouraged to create threads; library routines should
not create threads at all.

The introduction of the stream-entity construct in a programming
language is the \emph{first crucial addition} required by our steaming
model. In this model, kernel functions define computations while 
inter-stream-entity relations define precedence constraints in concurrent 
executions. Inter-stream-entity communications are purposely left out because
they are considered as implementation details; both distributed memory
methods (e.g. message passing) and shared memory methods are allowed.

At program control level, two more constructs may be desirable: one
for data parallelism and one for task parallelism. In the data parallelism
case, on the one hand, concurrent executions have a flat structure
in term of precedence constraint; some stream entities may have 
multiple instances at runtime, when each instance takes a portion
of a large stream data set. On the other hand, data partition
algorithms are often non-trivial. Therefore, a language construct, 
such as \texttt{map} in pMatlab~\cite{pmatlab}, will be useful for 
conveying the data partition information. Loops had often been used 
to do the job, but expressing parallel concept in sequential 
executions reduces program clarity.

In the task parallelism case, sometimes concurrent tasks are divided 
into groups for completely different functionalities, and have limited
synchronization points; sometimes increased security requirements,
such as those of a modern web browser~\cite{barth2008}, demand greater
isolation among the tasks. Thus, a task parallelism construct 
may be needed to define such parallel structures to allow special 
runtime scheduling and layout treatments.

The research community has been trying to invent new language 
constructs for the needs of multithreaded parallel computing,
such as in the cases of Cilk/Cilk++~\cite{cilk,cilk++} and Galois~\cite{kulk}.
The former emphasized the idea of separating designs for computation
from concerns for runtime load-balancing and scheduling; the
latter emphasized auto-parallelization of sequential programs 
with new language abstractions. Our language designs align well
with those interests, but are not handicapped by the reliance on
POSIX-like multithreading model.

\subsection{Compiler}
\label{sec-compiler}

The introduction of new language constructs such as that for stream
entity has two major impacts on compiler designs.

First, increased power in program analysis and parallelization. 
Stream entity defines a scope that can be bigger than loop nests,
and the scope is finite so is superior than intractable number 
of procedures that hinder previous practices in interprocedural 
analyses~\cite{allen,hall2005}. In the GeneSC model, a control 
construct for task parallelism as mentioned in Section~\ref{sec-lang} 
can guide code generation; a control construct for data parallelism can 
specify the number of concurrent streams, if programmers 
provide that information.

\begin{figure}[h]
\vspace{0.05cm}
{\centering \resizebox*{0.45\textwidth}{!}{\includegraphics{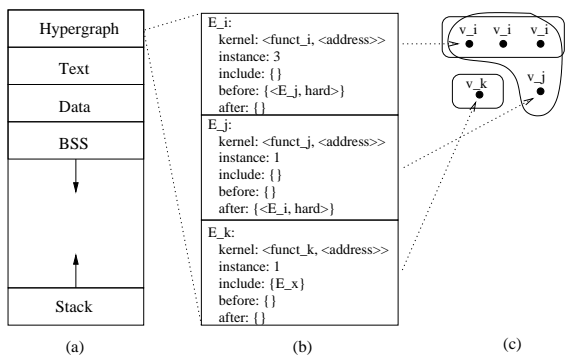}} \par}

\caption{\small Program virtual address space layout (a),
with hypothetical internal structures (b), and drawing of a hypergraph (c).}
\label{fig:layout}
\end{figure}

Second, added to compiler generated code thus a program's virtual address 
space layout, see
Figure~\ref{fig:layout}, is a new segment for storing hypergraphs
of inter-stream-entity relations, where the hypergraph vertices are 
stream entities, and the edges are sets of data-flow- or 
control-flow-connected stream entities; the size of an edge can be one.
The hypergraph information is the \emph{second crucial addition} 
enabled by the proposed stream computing model.

A hypergraph vertex, i.e. a stream entity, is designed as a non-distributive
execution unit, thus compilers (and later runtime schedulers) are allowed
to optimize computations through restructuring (e.g. flattening) the
initial hierarchical hypergraphs derived from user programs.

Architectural designs~\cite{exochi,pangaea} have been proposed to provide
much homogeneous programming environment on asymmetric or heterogeneous
multicore. Such works make mapping stream entities to asymmetric multicore
through compilers practically achievable. Depending on what the
underlying processor architecture is, just-in-time (JIT) compilation may
be employed to further optimize a program at runtime.

\subsection{Operating System}
\label{sec-os}

Program speedup technologies that take advantage of parallelism, such as 
instruction-level pipelines, data-level SIMD, and task-level 
hyperthreading~\cite{pattersonbook,ia64-32}, all have hardware assistants.
However, in the multicore case, the up-scaled, parallel processing
resource is exposed directly to programmers. Specifically, parallel task 
creation, mapping and scheduling to multiple cores become programmers' 
responsibility, with or without operating system involvement. Such arguable 
architectural defect or design trade-off may be regarded as the chief 
culprit for current difficulties in parallel programming on multicore 
processors. Architectural support for multi-thread (-block, -fiber,
-shred, -strand, etc.) programming, like those in~\cite{sequenzer,exochi}, 
would be helpful. Unfortunately, to our best knowledge, no such CPU production 
exists. To solve that problem, we propose software emulated, 
e.g. operating-system initiated, task pipelines where a task is a stream 
entity plus stream data. Such superscalar pipelines are the \emph{third 
crucial addition} enabled by the proposed streaming model. To illustrate 
the idea, we use the following shared memory model for more detailed 
discussions.

First, change how kernel creates and destroys a program's runtime 
image. For example, when the \texttt{exec} system calls are invoked,
a kernel thread, named micro scheduler to be distinguished from the 
process level scheduler, parses a program's hypergraphs information,
and decides the number of worker threads that form the superscalar
pipelines. Work-load information and power management instructions
may be available to the micro scheduler. Ideally, it will take those
information into consideration when creating worker threads. To 
reduce context switching\footnote{Context switching is a multiprogramming
strategy on single core. In multicore systems, the strategy may need
enhancement or extension.} overhead introduced by process level
scheduling, the micro scheduler dynamically changes the number of
worker threads.

\begin{figure}[h]
\vspace{0.05cm}
{\centering \resizebox*{0.45\textwidth}{!}{\includegraphics{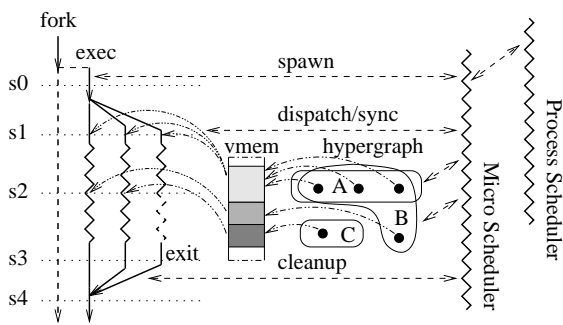}} \par}

\caption{\small A hypothetical case of operating system kernel initiated 
threads for stream computing.}
\label{fig:threads}
\end{figure}

Second, schedule a process' concurrent instruction blocks through 
work-stealing~\cite{cilk}. Although the micro scheduler speculates 
inter-stream-entity relations and dynamically control the number of
superscalar pipelines (i.e. worker threads), stream entities are
scheduled through work-stealing. Parsed and maybe further optimized 
hypergraphs information is saved as process context for reuse.
Since different operating systems have very different thread 
models~\cite{love,solaris}, we do not enforce how worker threads 
are implemented. Figure~\ref{fig:threads} shows a hypothetical 
case where the application process has three worker threads,
assisted at runtime by a micro scheduler.

Third, monitor virtual memory accesses and manage worker threads 
synchronizations based on an application's hypergraphs. Virtual 
memory access control is enforced through shadowing, e.g. locking 
a memory address so that only one worker thread can write to the 
address at a specific time; or coloring, i.e. making certain 
addresses accessible by only one of the worker threads at any 
time. When race conditions happen, block the contender thread(s) 
or issue a signal. One immediate question is: how can such 
fine-grained memory coloring be possible, if modern virtual
memory systems are page based? We are considering a dynamic version of  
memory overlays~\cite{levine} on top of virtual memory.
Memory overlays provide task-level program isolation.

Finally, let operating system kernel cleanup the worker threads
at process exit, and output tractable runtime states in case of
errors. For example, when system signals make a process abort,
hypergraphs information is dumped to a core file. A snapshot of
current running stream entities is useful but may not be 
possible in practice.

We've noticed a recent publication~\cite{torr} on architecture design
built on a concept called chunk execution, which was defined 
as a set of sequential instructions executed as one operation unit
with processor hardware assistance. Although it is still early
for us to assess the hardware design itself, the chunk execution idea
reaffirmed us the value of our task-level pipelines for multicore.

\section{Application}
\label{sec-app}

The modern web browser as a computing platform and a familiar but non-trivial
application is used to show the advantages of stream computing.

Besides plain text HTML contents, current and future web browsers
are expected to present to end users dynamic and rich media such as video,
3D images, virtual worlds, computer games, semantic web 
information, etc. Those complex contents require part, 
if not all, of the computations be done on user computers 
or smart hand held devices. For each different content format, 
the browser may run a virtual machine, e.g. a language interpreter, 
to process the content. There is also the need to isolate 
content handling tasks inside the browser for privacy and 
security reasons~\cite{barth2008, reis2009}. 

\begin{figure}[h]
{\centering \resizebox*{0.45\textwidth}{!}{\includegraphics{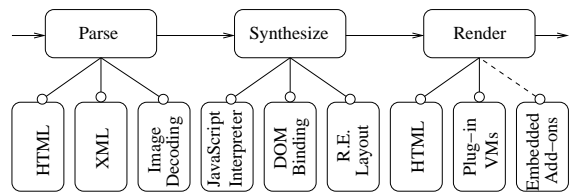}} \par}

\caption{\small Components in a simplified pipeline of the
rendering engine in a modern web browser.}
\label{fig:web}
\end{figure}

Figure~\ref{fig:web} shows a simplified pipeline of 
the rendering engine in a browser. Each of the three functional
units, namely parsing, synthesis, and rendering has multiple
sub-units of different capabilities, and the sub-units may be 
composed of different algorithms in implementation.

In this case, parallel programming using POSIX threads or OpenMP 
APIs would be hard. We see at least three problems with the 
user or compiler-initiated multithreading model: (1) Maintaining scalable 
number of threads while optimizing parallelization is difficult.
(2) Thread creation and destroy overheads may be significant, where 
threads are temporary objects. (3) Thread isolation is not guaranteed.

By using the proposed streaming model, computations are defined by
stream entities. Each component in Figure~\ref{fig:web} can be
built with stream entities through addition, concatenation, and
composition operations. At runtime, the micro scheduler defined 
in Section~\ref{sec-os} keeps a dynamically scalable number of
worker threads in a browser, and schedules stream entities as
execution units. Isolation is enforced through memory coloring
in the browser's virtual address space.
 
\section{Related Works}
\label{sec-related}

All directly related works have been
mentioned in the previous sections. What we discuss here are 
publications that may provide alternative solutions to various 
aspects of the problems we try to solve, or are noteworthy
references to earlier works. We skip publications on virtualization, 
threads scheduling and thread models, because they are highly 
specialized topics concerning implementations.

A school of analytical models, represented by the most recent Multi-BSP~\cite{mbsp}, provides theoretical abstractions that bridge parallel algorithm 
designs with multicore architectures. One common trait of such models
is their emphasize on portable performance. However, such motivation
may not ensure engineering success, as there are more imminent concerns,
such as indeterminacy, synchronization overhead, and scalability, that 
highlight the weaknesses of current parallel programming tools. In the 
Multi-BSP case, the model should be useful in analyzing a given parallel 
computing system in some multicore architectures, but the model does 
not provide enough engineering guidance in designing such a system. 
It is not surprising that the Multi-BSP model's hierarchically nested 
component structures resemble the stream entity hypergraphs in our model, 
but our stream entities have richer engineering implications.

Someone has suggested that the GeneSC model that we promote was a rediscovery
of Kahn's Process Networks~\cite{kahn74, kahn77}. But that's a misunderstanding,
if one can ignore the superficial resemblance in our use of certain
terminologies. The simple language Kahn defined and discussed is both
synchronous and deterministic, due to the language's blocking primitives
such as {}``wait'' and {}``send'', which build the language's FIFO 
queuing model for communication. The Kahn model suits concurrent tasks, 
but has little emphasis on high performance computation.
We argue that a concurrent computing model should convey intrinsic data 
and task precedence constraints but should not sacrifice asynchronism
when it is possible. The flexibility provided by asynchronism is 
essential for achieving high-performance through out-of-order execution,
thus is needed by a broader range of applications.

In his thesis on StreamIt~\cite{thies2009}, Thies summarized languages
from Kahn's Process Networks to synchronous dataflow~\cite{ealee3} and CSP 
(Communicating Sequential Processes)~\cite{csp2004}. Such languages 
have a common feature that is the pipe-lining of concurrent task-executions.
StreamIt also experimented on the idea of stream graph that allows compiler
optimizations. Our main reservation with such languages is their weak power
in modeling data parallelism.

Jade~\cite{jade} represents an important experiment on parallel programming
language design. Jade preserves the serial semantics of a 
program, implicitly exploits task parallelism, and moves data closer 
to processors. However, if it was designed for a smooth transition
from sequential programming to parallel programming, Jade's dependence on
its type system and explicit object-access control still exposes 
low-level synchronizations to programmers. While arguably such low-level
controls provide programming flexibility, not all the details are
essential to algorithm design. On the other hand, we might consider that
Jade lacks the power for defining concurrent execution structures in a 
larger program-scope.

Another
parallel programming language that has limited adoption in the academia
is UPC~\cite{upc}. UPC is another multithreading language that extends C.
The two most noteworthy features of UPC include the partitioned global
address space model and the memory consistence model, where the latter
is implemented using explicit access controls to shared objects. Again,
the weakness of UPC might be its explicit synchronizations exposed to 
programmers.

Chapel~\cite{chapel} is a complex parallel programming language that
supports global level data and control abstractions, architecture-aware
locality mapping, object-oriented programming and generic programming.
What impressed us most are the language's global views on data and control,
and its ability to map the global views onto data/control affinity
abstractions called locales. Though we do not have experience with 
Chapel programming, by studying its ancestor ZPL~\cite{snyder2007}, we
see one motivation that emphasizes both performance
and portability on various parallel architectures. For that purpose, 
ZPL and Chapel may strike for a balance between abstraction and
expressiveness. The emphasis of our GeneSC
model, however, is (high-level) concurrency structures in problems'
solutions. Therefore, we can rely on extensions to existing
languages to achieve concurrency structure abstraction, and 
delegate performance concerns (e.g. the mapping to parallel architectures)
to languages and compilers that implement the kernel functions 
encapsulated by a stream entity.

We've mentioned Cilk++ earlier but consider that it deserves more
attention. First, Cilk++ is interesting because of its theoretical
foundation and its internal handling of both task-parallelism and
data-parallelism. Cilk++ has a provable threading model that can
be analyzed using DAG (Directed Acyclic Graph). Conceptually, a
Cilk++ strand is like a stream entity of our model, though Cilk++
does not have an explicit construct for strand. Second, Cilk++
uses only two language constructs, namely \texttt{cilk\_spawn} and
\texttt{cilk\_for}, to define task parallelism and data parallelism
at language level, though Cilk++ language constructs have scope
restrictions that make C++/Cilk++ mixed-language programming
cumbersome~\cite{cilk++}. Finally, Cilk++ has restricted but
innovative debugging model and hyper-objects support.

In a survey~\cite{mccool} on parallel computing languages for multicore 
architectures, McCool compared and contrasted existing programming 
language models and computing paradigms. One shared interest is about 
stream computing, particularly the {}``SPMD stream processing model'', 
implemented on the RapidMind platform~\cite{rapidmind}. Three of the 
platform's major contributions may be summarized as follows. 
First, at language level, RapidMind adopts a type-system approach 
inlining with Jade, CUDA and OpenCL; but the type system is much
simpler and exposes no synchronization nor access control details.
Second, at runtime level, RapidMind schedules and load-balances 
the executions of concurrent {}``program objects'' through backend
(runtime) supports, which identify and parallelize the program objects.
Noticeably, a program object has similar purpose as our stream entity 
but lacks of precedence constraints. Third, the RapidMind platform
supports heterogeneous multicores. One needs to be reminded, however, 
that while it does not relying on conventional threads, RapidMind's
implementation of the SPMD stream processing model limits the  
platform to applications that show data parallelism.

In~\cite{lithe},
the authors proposed a close-to-metal substrate called {}``lithe'',
which uses a primitive named {}``hart'' and its context to
carry out application-level threads in multicore processors.
Lithe also has an interface layer that allows processor 
allocation and thread scheduling through runtime systems. The 
motivation was to control hardware resources 
subscription and to support parallel library compositions. As an 
alternative OS-level thread implementation, Lithe has been used 
to run existing multi-threading libraries such as OpenMP.

Bl\"{a}ser in~\cite{blaser} reported a
component-based language and operating system that supported concurrent
and structured computing. In that system, a component encapsulates
data and computing, which are defined as services to be produced by
one component and consumed by another. The emphasis on relations among
components is like ours, except that communications among
related components use explicit message passing.

Finally, the GeneSC model is for concurrent 
computing at instruction-block level. Hardware-dependent considerations,
such as supports to manycore and heterogeneous multicore,
will rely on implementations, particularly those at runtime and 
operating system levels, such as those in~\cite{multikernel,tessellation,helios}.

\section{Concluding Remarks}
\label{sec-conclude}

In this paper, we discuss a generalized stream computing model, 
known as the GeneSC model, for concurrent computing. This model is 
not about yet another programming technique or a programming paradigm, 
but an abstract design guidance for parallel programming languages, 
compilers, runtime and/or operating systems for multicore. We 
highlight the benefits of the model through operational analysis, 
and demonstrate its power through an enabled programming framework
and example design considerations that support the framework. 
Specifically, the GeneSC model brings in a non-distributive 
execution unit called stream entity, and uses it as the foundation 
for designs at programming language level, compiler level, 
and operating system level. Besides stream entity, two other crucial
additions unlike traditional practices are also introduced: a
new hypergraph section in a program's runtime layout, and operating
system kernel initiated threads that form stream-entity/stream-data
pipelines to multicore processors. The significance of the GeneSC
model is that, by harnessing concurrency information expressed in 
algorithms and by applying the information to runtime scheduling, 
the model provides tractable, concurrent operations; and may serve
as a natural extension of the von Neumann model for parallel 
computing architectures.

Our GeneSC model facilitates a high-level abstraction of concurrency 
in computing, yet does not incur a deep learning curve because the 
kernel of a stream entity is the encapsulation of a single-/multiple-procedure 
function that can be implemented in existing sequential languages.
By delegating many of the tough issues (e.g. synchronization 
and task scheduling) in concurrent computing to platform software such
as compilers and operating systems, programmers can focus more on 
analyzing their application problems, designing algorithms, and 
defining concurrency structures of the solutions---a shift from 
machine-oriented programming to application-oriented programming;
at the meantime, sequential programming develops smoothly into 
concurrent programming.

\vspace{0.4cm}

\textbf{\emph{Acknowledgments.}} The author learned of the existence 
of hypergraphs through professor Vitaly Voloshin of the Troy 
State University, USA. An anonymous reader of the first draft
of this paper brought up Kahn's works, which led to the
comparison between Kahn's model and ours. The author once questioned
professors Peter J. Denning and Jack B. Dennis about their
discussion of determinate computation in an article published in CACM, 
Vol. 53, No. 6; the two professors' kind replies further clarified this 
author's use of certain terminologies.

\end{document}